\newcommand{\beq}{\begin{equation}}
\newcommand{\eeq}{\end{equation}}

\documentclass[%
 amsmath,amssymb,
]{revtex4-1}

\usepackage{color, graphicx}
\usepackage{dcolumn}
\usepackage{bm}
\usepackage{amssymb}
\usepackage{amsmath}
\usepackage{latexsym}
\usepackage{amsfonts}
\usepackage{amsmath}
\usepackage{multirow}
\usepackage{ifthen}

\begin{document}


 \title{Transparency robust to disorder in a periodic array of Helmholtz resonators}

\author{O. Richoux}
\affiliation{LUNAM Universit\'e, Universit\'e du Maine, CNRS, LAUM, UMR 6613, Av. O. Messiaen, 72085 Le Mans, France}
\author{A. Maurel}
\affiliation{Institut Langevin, LOA, CNRS 7587, ESPCI, 1 rue Jussieu, 75005 Paris, France}
\author{V. Pagneux}
\affiliation{LUNAM Universit\'e, Universit\'e du Maine, CNRS, LAUM, UMR 6613, Av. O. Messiaen, 72085 Le Mans, France}

\date{\today}

\begin{abstract}

In this paper, the influence of disorder on 1D periodic lattice of resonant scatterers is inspected. These later have multiple resonance frequencies which produce bandgap in the transmission spectrum. One peculiarity of the presented system is that it is chosen with a nearly perfect overlap between the Bragg and the second hybridization bandgaps. In the case of a perfect ordered lattice, and around this overlap, this produces  a narrow transparency  band within a large second bandgap. As expected, the effect of the disorder is generally to increase the width of the bandgaps. Nevertheless, the transparency band  appears to be robust with respect to an increase of the disorder. In this paper, we study this effect by means of experimental investigations and numerical simulations.
 \end{abstract}
 
\maketitle

\section{Introduction}
Phononic crystals have experienced an increasing interest in recent years because of their potential applications to acoustic filters \cite{Romero}, the control of vibration isolation \cite{Hussein}, noise suppression, and the possibility of building new transducers \cite{Wu}; for a review see \cite{Page}. It is thus of interest to understand which properties of such structures are sensitive to inherent imperfections in their design and which are not. Besides, one can also address the question of whether or not the disorder can make new interesting properties to appear.

It is usual to characterize a random medium in terms of an effective -homogeneous- medium. For random perturbation of homogeneous free space one finds that the dispersion relation $K(\omega)$ departs from the dispersion relation $k(\omega)$ in free space without disorder, and the imaginary part of the effective wavenumber $K$ indicates how much the opacity due to disorder is important \cite{lintonmartin}. 
 In the case of photonic or phononic crystals, the band structure of the unperturbed medium is more complicated, with a wavenumber $Q$ of the Bloch Floquet mode being either purely real (pass band) or complex (stop band). The addition of disorder modifies  the band structures of these periodic-on-average systems  \cite{maradudin,Deych98, Han2008, Maurel2008, Maurel2010,Izrailev2012,Maurel2013}, and generally, produces an increase in the band gap width \cite{Chang2003}. Among periodic media, the case of periodic arrays of resonant scatterers is very attractive since the resonances inherent to the individual scatterers produce strong modifications of the wave propagation; owing to these modifications in the wave properties may help to design materials with unusual properties. Such arrays present band gaps around the resonance frequencies of an individual scatterer. Because periodically located, Bragg resonances are also produced, resulting in a complex  band gap structure. Overlapping two types of gaps, a resonant scatterer gap and a Bragg gap, have been shown to produce interesting phenomena, as the creation of a super wide and strongly attenuating band gap used for structure isolation \cite{Croenne2011,Xiao2011,Sugimoto1995,Bradley1994} and slow wave application \cite{Theocharis2014}. 

In this paper,  we consider the propagation of an acoustic  wave in  a periodic  array of Helmholtz resonators  connected to a duct in the plane wave regime (low frequency regime with one propagating mode in the duct). 
The corresponding model describes the  1D propagation of the  pressure field $p(x)$ through resonant point scatterers (Kronig- Penney system) \cite{Olivier,Richoux2009}
\beq
p''+k^2p =  \sum_n V_n(k)\delta(x-nd)p(x),
\label{WEpotential}
\eeq
where $d$ is the periodicity of the array and where  $V_n(k)$ encapsulates the effect of the $n$th resonator of the array.  
The disorder is introduced by varying the volume of the Helmholtz resonators. When an overlap between a Bragg gap and a resonant bandgap is produced, a narrow transparency band appears within the resulting large bandgap.  Unexpectedly, we found that this transparency band is robust with respect to the disorder. 
Indeed, first, for small disorder, the transmission decreases; but increasing further the disorder induces an increase in the transmission.
 We have carried out experiments whose results show qualitatively this behavior. 
 To get further informations, with a broader range of the disorder parameter, numerical calculations are shown, that confirm the transparency induced by disorder. The paper is organized as follow: in Part \ref{part1},  the 1D model and the CPA result for the randomly perturbed system are discussed. The  experimental results are presented in Part \ref{part2}, and this is completed by numerical calculations, in Part \ref{part3}. Finally, a discussion is proposed in Part \ref{part4}.


\section{Propagation in 1D periodic and perturbed HR array}
\label{part1}

At low frequencies, when only one mode can propagates in the duct, the propagation of acoustic waves in an  array of Helmholtz resonators periodically located with spacing $d$  (Fig. \ref{Fig_exp_setup}) can be described by
\beq
p''+k^2p =  \sum_j V_j(k)\delta(x-jd)p(x),
\label{WEpotential}
\eeq
where $p$ is the pressure field and $k=\omega/c_0$ (the time dependance $e^{-i\omega t}$ is omitted, $\omega$ is the angular frequency and $c_0$ the sound velocity in free space).
The potential is 
\beq
V_j(k)= - \frac{s}{S_w} \;k_n\; \frac{ \sin k_n\ell \cos k_c L+\alpha \cos k_n \ell \sin k_c L}{\cos k_n \ell \cos k_c L - \alpha \sin k_n \ell \sin k_cL},
\label{potentialV}
\eeq
with $\alpha= Sk_c/(sk_n)$, where $S_w, S, s$ are the area of the main waveguide,  of the cavity and of the neck, respectively. 
$\ell$ and $L$ denote the length of the neck and of the cavity respectively (see Fig. \ref{Fig_exp_setup}(b)). The wavenumbers are $k_m=k\left[ 1+ \beta \delta/R_m\right]$, with $m=w,c,n$ (waveguide, cavity and neck respectivelly) and $R_m$ the corresponding radius, with $\beta=\left[1+(\gamma-1)Pr^{-1/2}\right](1+i)/\sqrt{2}$ and $\delta=\sqrt{\nu/\omega}$ the viscous boundary layer depth ($\nu$ the cinematic viscosity).
The term proportional to $\beta$ in the wavenumber $k$ is a good model of the viscous and thermal attenuation of sound in the duct. We can notice that, with $s \ll S_w$, the strength of the Helmholtz scatterer is small except at resonances. 
Approximating $k_n$ and $k_c$ by $k$, and thus omitting the attenuation, these cavity resonances correspond to a vanishing term $D(k)\equiv \cos k\ell \cos kL - \alpha \sin k\ell \sin kL$, and they are of two types: i) the typical Helmholtz  resonance occurring at low frequency,  say for  $k\ell\to 0$ close to $k_H=1/\sqrt{\alpha \ell L}$ and ii) the resonances in the cavity (hereafter referred as volume resonances), near $kL=n\pi$. For instance, for $n=1$,
\beq
k_V L=\pi+\frac{1}{\alpha \tan(\pi \ell/L)}.
\eeq
For a single resonator, these resonances produce a vanishing transmission. 
When the resonators are organized in a perfect periodic array, 
band gaps are created around the resonance frequencies, according to Bloch Floquet wavenumber $Q$ becoming purely imaginary, $Q$ being given by
\beq
\cos Qd= \cos kd +  \frac{V}{2k} \sin kd. 
\eeq
When disorder is introduced in the volume cavity by changing the length $L_n$ of the n$^{th}$ cavity, $L_n = L (1+\epsilon_n)$ and $\epsilon_n \in [-\epsilon/2;\epsilon/2]$, it is possible to predict the new Bloch Floquet $K$ using CPA approach  
\cite{Maurel2010}
\beq
\cos Kd= \cos kd +  \frac{\langle V\rangle}{2k} \sin kd,
\eeq
where $\langle .\rangle$ denotes the ensemble average for all realizations of the $\left\{\epsilon_n\right\}_n$-values. 
The resulting transmission coefficient is
\beq
T_N=\frac{e^{ikd}-{\cal B}^2e^{-ikd}}{e^{ikd-iKNd}-{\cal B}^2e^{-ikd+iKNd}},
\label{TN_CPA}
\eeq
where we have written $p(x\geq Nd)=T_Ne^{ik(x-Nd)}$  (the incident wave is  $e^{ik x}$) and
with 
\beq
{\cal B}\equiv  \frac{1-e^{i(k-K)d}}{1-e^{i(k+K)d}}.
\eeq
Obviously, the above results obtained from CPA recover the perfect periodic case when  $\epsilon=0$.

In the following, we present the experimental set-up to realize the lattice of Helmholtz resonators. Comparisons between the measured transmission and the above CPA- result, Eq. \eqref{TN_CPA} is presented.

  \section{Experimental results}
  \label{part2}
\subsection{Experimental set-up}

\begin{figure}[h!]
\begin{center}
\includegraphics[width=8.cm,clip]{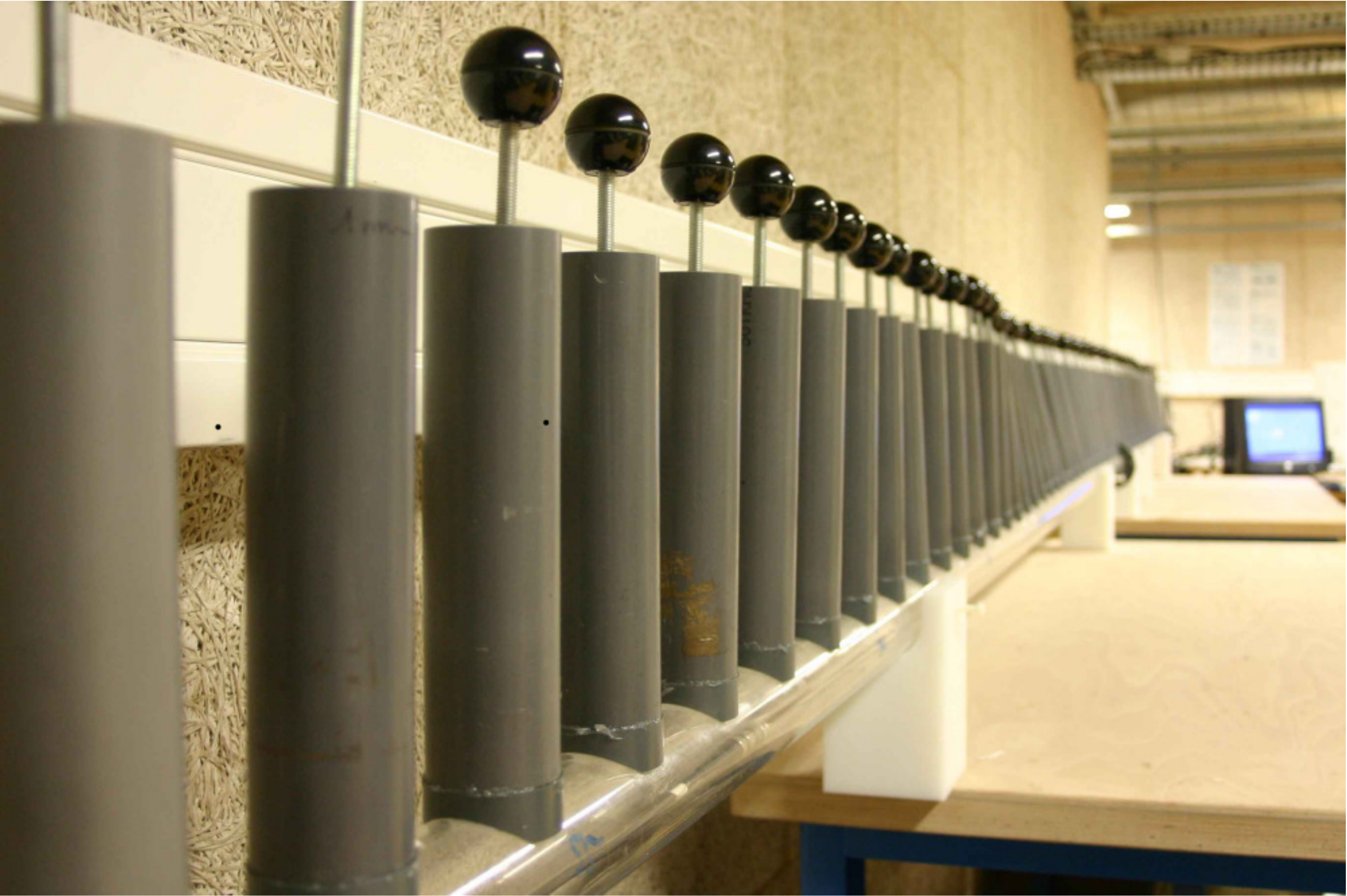}
\includegraphics[width=8.cm,clip]{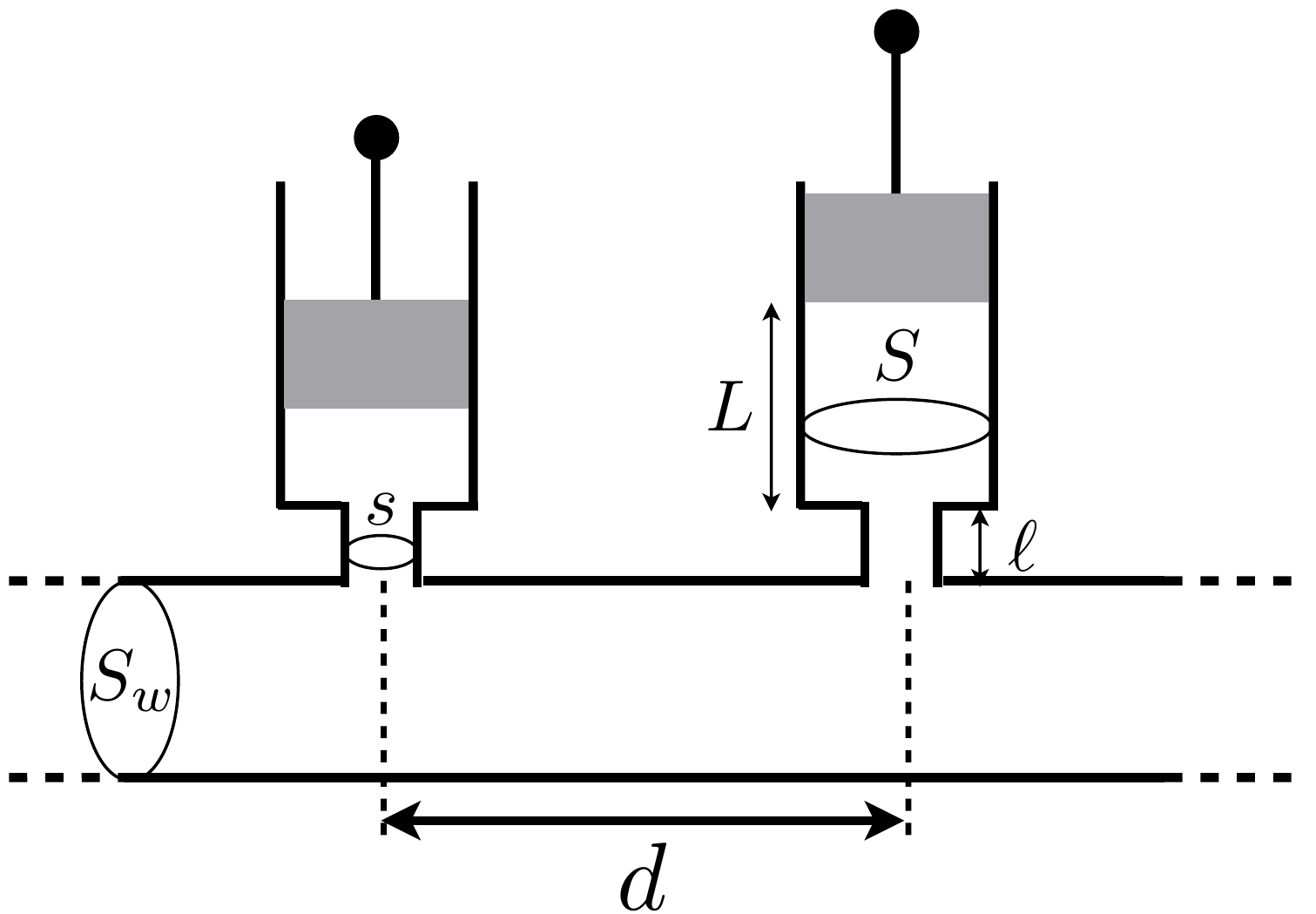}
\end{center}
\caption{Picture of the experimental set up (left panel). Shematic of the experimental setup (right panel).}
\label{Fig_exp_setup}
\end{figure}

The experimental set-up (fig. \ref{Fig_exp_setup}) consists in a $8$ m long cylindrical waveguide with an inner area $S_w=2\times10^{-3}$ m$^2$ and a $0.5$ cm thick wall. This waveguide is connected to an array of $N=60$ Helmholtz resonators periodically distributed, with inter-distance $d=0.1$ m. Each resonator is composed by a neck (cylindrical tube with an inner area $s=7.85\times10^{-5}$ m$^2$ and a length $\ell = 2$ cm) and by a cavity with  variable length.  The  cavity is a cylindrical tube with an inner area $S=1.4\times10^{-3}$ m$^2$ and a maximum length $L_{max}=16.5$ cm, see fig. \ref{Fig_exp_setup}(b).

The sound source is connected to the input of the main tube. 
The source is embedded in an impedance sensor \cite{MinutePub}, able to calculate the input impedance of the lattice $Z$, defined as the ratio of the acoustic pressure $p$ and the acoustic flow $u$ (the product of the velocity by the area cross section) at the entrance of the lattice, as described in \cite{Macaluso2011}. 
This allows  to get the reflection coefficient  $R$ defined as $p=(1+R)p^+$ owing to $u=u^++u^-$ with $u^+=p^+/Z_w$, $u^-=-p^-/Z_w$ , where the index $+$ and $-$ denote the parts of the quantity associated to right- and left-going waves:
\beq
R=\frac{Z-Z_w}{Z+Z_w}.
\eeq

At the output, an anechoic termination made of a $10$ m long waveguide partially filled with porous plastic foam suppresses the back propagative waves. This ensures the output impedance to be close to the characteristic impedance  $Z_w=\rho c /S_w$. Finally, a microphone is used to measure the pressure $p_e$ at the end of the lattice. 

Using line matrix theory, $(p,u)$ and $(p_e,u_e)$ are linked by the transfer matrix through
\beq
\left(
\begin{array}{c}
p\\ u
\end{array}
\right)
=\left(
\begin{array}{cc}
A & B \\ C & D\end{array}
\right)
\left(
\begin{array}{c}
p_e\\ u_e 
\end{array}
\right)
\eeq
with $ p=Zu$ ($Z$ being measured) and $u_e= p_e/Z_w$ (the acoustic flow is deduced from $p_e$ because of the anechoic termination). Then, the transmission coefficient $T$ defined as 
$p_e=Tp^+$, is calculated using that $u=(1+R)p^+/Z$ by definition of $R$ and from above, $u=[C+D/Z_w]p_e=[C+D/Z_w]Tp^+$, from which 
\beq
T=\frac{2 Z_t}{Z+Z_w},
\eeq
where $Z_t\equiv [C+D/Z_w]^{-1}$ is deduced from the measured $(p_e, u)$-values.

When considered, the disorder in the lattice is introduced through the variable lengths $L_n$, $n=0,\dots, N$ of the cavities, and $L_n = L (1+\epsilon_n)$ is used with a normal distribution of $\epsilon$ being chosen for each realization and for each resonator cavity, with $\epsilon_n \in [-\epsilon/2;\epsilon/2]$, resulting in a variable scattering strength,  $V_n$ in Eq. \eqref{potentialV}.

The transmission coefficients are measured for ten different distributions with same standard deviation, and the mean value $\langle T\rangle$ is taken.

\subsection{Experimental observations}

\begin{figure}[h!]
\begin{center}
\includegraphics[width=12.cm,clip]{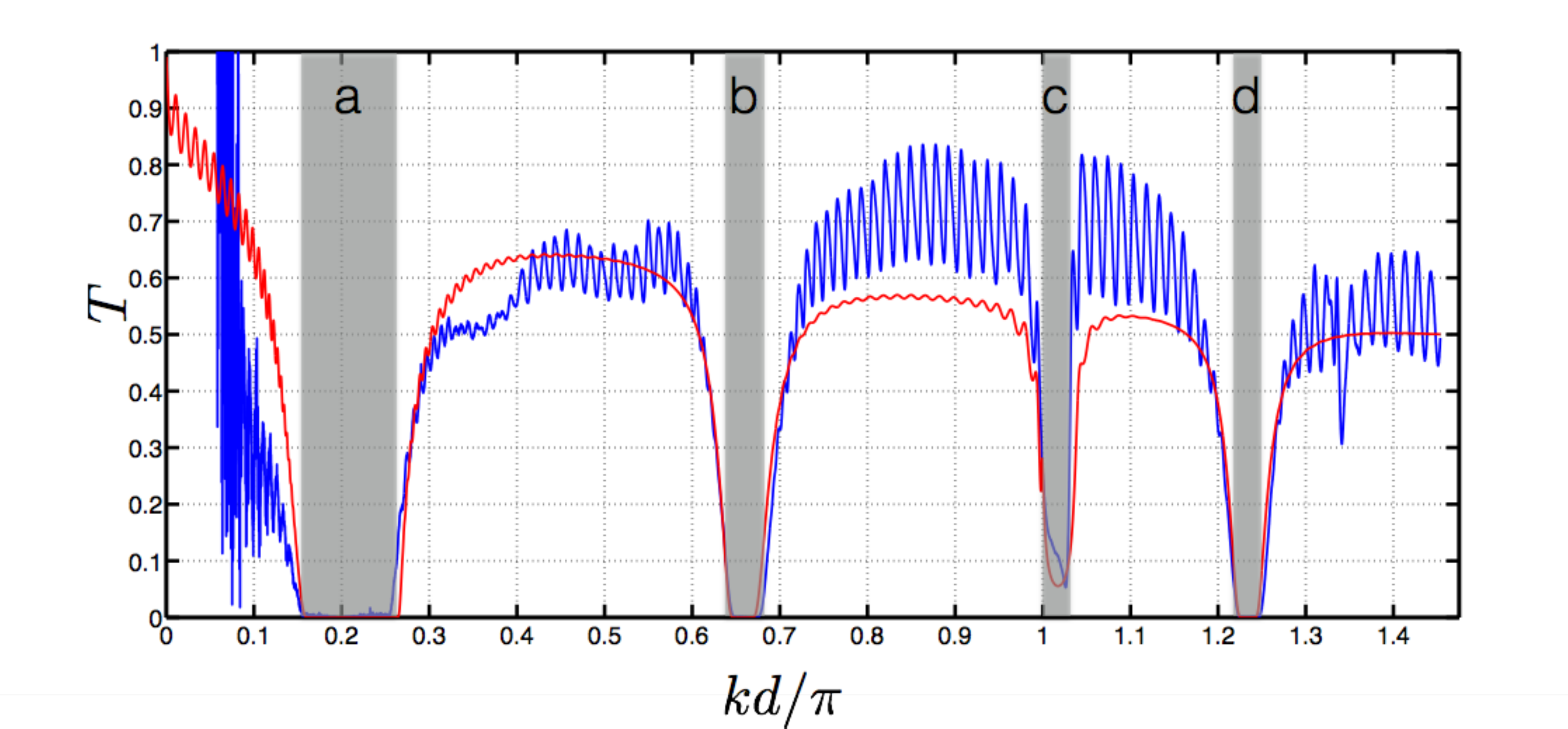}
\end{center}
\caption{\label{Fig_ord} Transmission coefficient for an ordered lattice with a cavity length $L=0.165$ m  and lattice spacing $d=0.1$ m. Blue line corresponds to the experimental measurement and red line to the analytical prediction, Eq. \eqref{TN_CPA}.}
\end{figure}

The transmission coefficient $T$ in the perfect periodic case is presented in the Fig. \ref{Fig_ord} for a cavity length  $L=0.165$ m. Four  bandgaps are visible : The first (labeled a) at low frequency is associated to the Helmholtz resonance ($k_H$) for $kd/\pi \in [0.15;\ 0.25]$, corresponding to frequency in $[300;\; 450]$ Hz. Two other band gaps (labeled b and d)  are associated to the two first volume 
resonances  ($kL$ close to $\pi$ and $2\pi$); these are for $kd/\pi$ in $[0.64 ;\; 0.68 ]$, $[1.22;\; 1.24]$ (corresponding frequency ranges $[1110;\;1170]$ Hz and $[2100;\;2150]$ Hz).
These three band gaps associated to resonances of the scatterer are  often referred as hybridization band gaps \cite{Croenne2011}.
Finally, the band gap labeled c is associated to the Bragg resonance, for $kd/\pi \in [1 ;\; 1.03 ]$, (frequency range $[1700;\; 1800]$ Hz). This band structure has been described in details, including non linear aspects, in \cite{Olivier,Richoux2007, Richoux2009}.
 The comparison between the experimental (blue line) and the analytical expression, Eq. \eqref{TN_CPA}, (black line) shows a good agreement. The discrepancy in the low frequency regime may be attributable to the bad quality of the source in this frequency range. 

 Finally, the strong peaks appearing in the experimental measurements  are due to the imperfection in the anechoic termination, resulting in interferences between forward and  backward waves in the main tube. 

\begin{figure}[h!]
\begin{center}
\includegraphics[width=12.cm,clip]{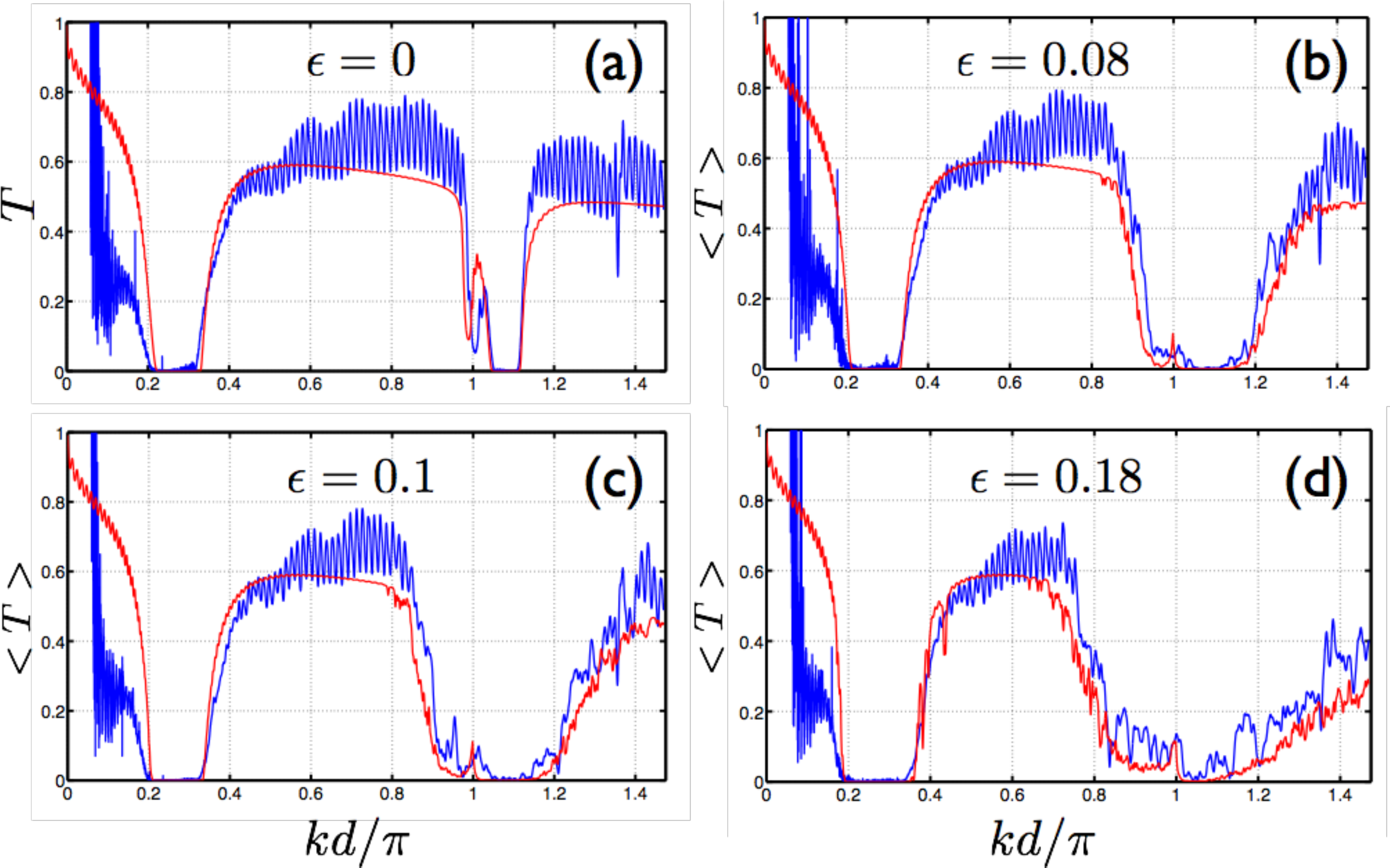}
\end{center}
\caption{\label{Fig_desordre} (a) Transmission coefficient of an ordered lattice for a cavity length  $L=0.1$ m and lattice spacing $d=0.1$ m. (b) Mean value of the transmission coefficient for a disordered lattice with $\epsilon=0.08$. (c) Mean value of the transmission coefficient for a disordered lattice with  $\epsilon=0.1$. (c) Mean value of the transmission coefficient for a disordered lattice with  $\epsilon=0.18$. The blue line corresponds to the experimental case obtained with $10$ averages and the red line to our analytical prediction with $100$ averages (except for (a) without disorder).}
\label{Fig_exp_result1}
\end{figure}

The Fig. \ref{Fig_desordre}(a) shows the transmission in the perfect periodic case for $L =0.1$ m. With $L=d$, the volume resonance  $k_V$ , with $k_V\sim \pi/L$,  and the Bragg frequency $k_B=\pi/d$ are very close, resulting in an almost perfect overlap of the two corresponding band gaps, previously labeled b and c, visible here in the range $kd/\pi \in [0.98;\; 1.12]$ (frequency range $[1600;\; 1800]$ Hz). The first band gap, associated to the Helmholtz resonance $k_H$ is almost non affected by the change in $L$ while the volume resonance with $k_VL\simeq 2\pi$ (previously labeled d) is sent to higher frequency (not visible in our plot). 
A noticeable feature is  the existence of a small transparency band inside the large stop band near $kd=\pi$, a feature already observed in other system where such overlapping is realized \cite{Sugimoto1995,Bradley1994,Theocharis2014}. This feature, in addition to the main behavior of $T$, is accurately captured by our analytical expression, Eq. \eqref{TN_CPA}, in the perfect periodic case, thus with constant unpertubed potential $V$ (and $K=Q$).

We now consider several amplitudes $\epsilon$ of  disorder in the scattering strength of the resonators, as previously described. 
The measured transmission coefficients $|\langle T\rangle| $ are reported in Fig. \ref{Fig_exp_result1}(b,c,d) for  respectively $\epsilon=0.08, \; 0.1, \; 0.18$. 

As expected, the more visible effect of the disorder is to strengthen the opacity of the media. This is associated to the fact that the wavenumber $K$ of the effective Bloch mode acquires an imaginary part due to the disorder (in addition to the attenuation) in the ex-pass bands of the perfect ordered case. In counterpart, in the ex-stop bands of the perfect ordered case,  the 
imaginary part of the wavenumber decreases, resulting in an increase of the transmission \cite{maradudin}.

In the second stop band, an interesting  behavior can be noticed, although very qualitative at this stage: 
 inside the second band gap, around $kd=\pi$, the small transparency band remains visible, since we observe a 
 peak of transmission robust to disorder. This trend is confirmed by the analytical model (red curves on Fig \ref{Fig_exp_result1}).
 
In the following section, we use numerical calculations to get further insights  on this  induced transparency near $kd=\pi$.

\section{Numerical inspection of the induced transparency}
\label{part3}

We now present results from numerical experiments  of the propagation in the array of Helmholtz resonators. This is done by solving Eq. \eqref{WEpotential}, with variable ${V}_n$ values. The disorder is introduced by using  $L_n=L(1+\epsilon_n)$  in Eq. \eqref{potentialV}. 
To calculate $p(x)$, we implement a method based on the impedance, as describe in  \cite{Maurel2008}.
For each frequency, $N=10^4$ realizations of the disorder with same amplitude $\epsilon$ are performed. The effective transmission $\langle T\rangle$ is calculated by averaging the transmission coefficients $\langle T\rangle=1/N \sum T_n$, where the $T_n$ are the transmission coefficients for each realization. 

\begin{figure}[h!]
\begin{center}
\includegraphics[width=12.cm,clip]{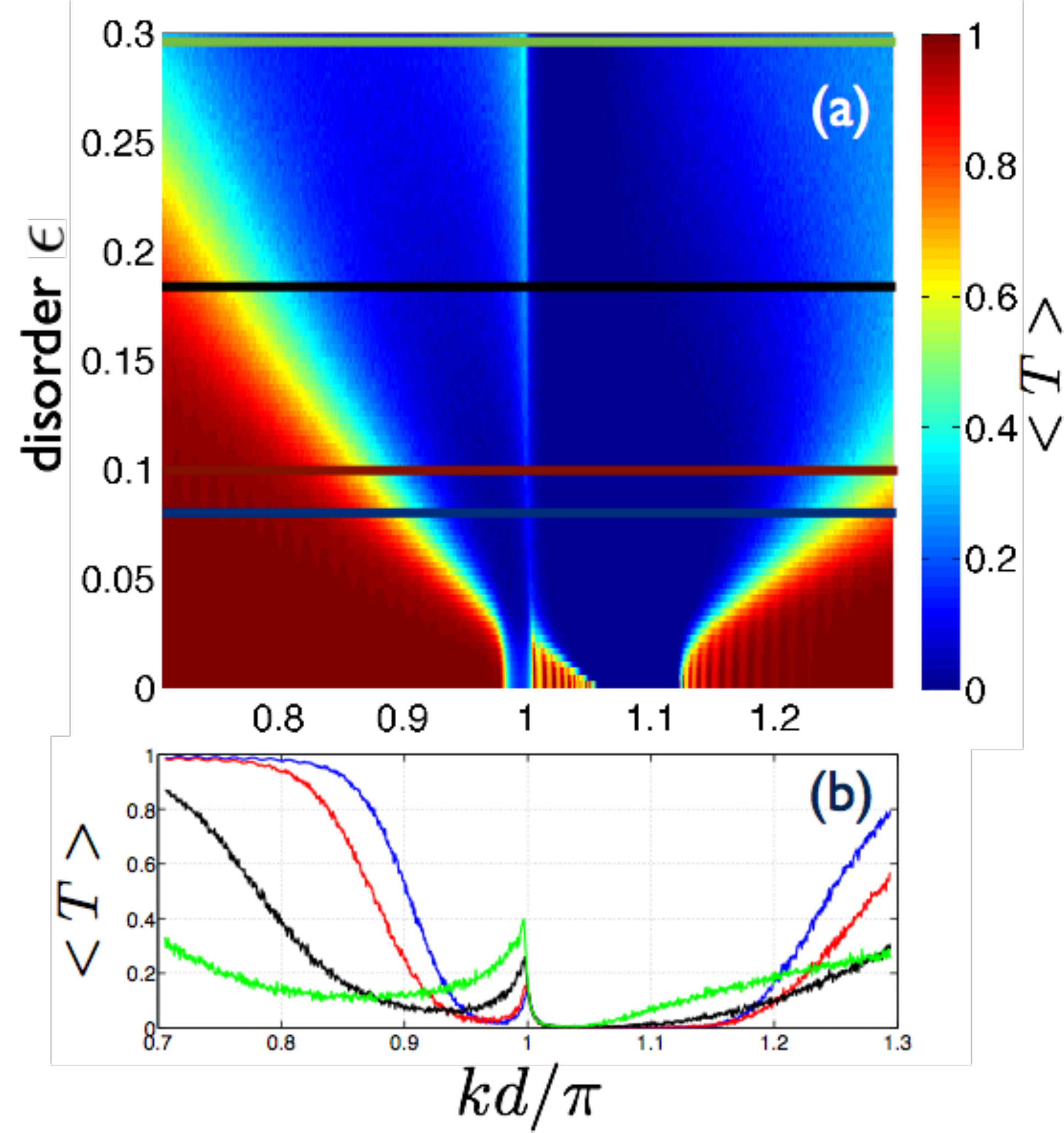}
\end{center}
\caption{(a) Mean value of the transmission coefficient in function of the disorder. (b) Mean value of the transmission coefficient for $\epsilon = 0.08$ (blue), $\epsilon=0.1$ (red), $\epsilon=0.18$ (black) and $\epsilon=0.3$ (green).}
\label{Fig_main_num}
\end{figure}

The main result is presented in Figs. \ref{Fig_main_num}. In Fig. \ref{Fig_main_num}(a), $|\langle T\rangle |$ is shown in a 2D plot as a function of $kd$ and $\epsilon$ and Fig. \ref{Fig_main_num}(b) shows several transmission curves for given $\epsilon$-values. 
Clearly, with $10^4$ realizations of the disordered, the averaged systems have converged.  
The transparency robust to disorder  is quantitatively confirmed: For the largest values of disorder, the transmission near $kd=\pi$ increases with the disorder.

\section{Discussion}
\label{part4}

Robustness of transparency to disorder could appear counterintuitive with regards to the usual influence of disorder in wave propagation. Indeed, the presence of disorder is known to break the wave propagation and to avoid any transmission. In this study, robustness of transparency is the result of the mixing of two different physical phenomena : (1) the non-exact overlap of the Bragg and hybridization band gaps which generates, in the periodic case, a narrow passband located inside a band gap and (2) the presence of disorder on potential which prevents the wave propagation inside the media. In the periodic case, one of the edges of transparency band due to overlap is located at $kd/\pi = 1$ which corresponds to the Bragg frequency \cite{Theocharis2014,Sugimoto1995}. Because disorder is injected in the potential (resonance frequency of the Helmholtz cavity), the edge of the transparency band is not affected \cite{Maurel2013}. As a consequence, the narrow passband is very sensitive to disorder and disappears from its upper edge remaining the lower edge unchanged and creating a peak of transparence for $kd/\pi = 1$. On the contrary, with no overlapping, the Bragg bandgap increases with disorder only from its upper edge. The lower edge belongs to a passband. In this case, there is no peak of transparency. This configuration can be used to filtering applications by tuning very narrow filter by injection of disorder in the system.

\section{Conclusion}
In this paper, we reported an experimental and numerical characterization of a periodic-on-average disordered system.
The usual  widening of the band gaps of disordered arrays is observed. 
On the other hand, when nearly perfect overlap between the Bragg and the scatterer resonance frequencies is realized, 
evidence of robust transparency has been shown.

\vspace{0.5cm}
\noindent
{\bf Acknowledgments}. This study has been supported by the Agence Nationale de la Recherche through the grant ANR ProCoMedia, project ANR-10-INTB-0914.
V.P. thanks the support of Agence Nationale de la Recherche 
through the grant ANR Platon, project ANR-12-BS09-0003-03.


\begin{thebibliography}{9}
\bibitem{Romero} 
V. Romero-Garcia, C. Lagarrigue, J.-P. Groby, O. Richoux and  V. Tournat, J. Phys. D \textbf{46}, 305108 (2013).
\bibitem{Hussein} 
M.I. Hussein, K. Hamza, G.M. Hulbert and K. Saitou, Waves in Random and Complex Media {\bf 17}, 491 (2007).
\bibitem{Wu}
 T.Z. Wu and S.T. Chen, IEEE Trans. Microwave Th. and Tech. {\bf54}, 3398 (2006).
\bibitem{Page}
 J.H. Page, A. Sukhovich, S. Yang, M.L. Cowan, F. Van Der Biest, A. Tourin, M. Fink, Z. Liu, C.T. Chan and P. Sheng, Phys. Stat. Sol. B {\bf 241}, 3454 (2004).
  
\bibitem{lintonmartin} 
C. M.  Linton and P.A. Martin,  J. Acoust. Soc. Am. {\bf 117}(6), 3413-3423 (2005).


  \bibitem{maradudin}
V. D. Freilikher, B. A. Liansky, I. V. Yurkevich, A. A. Maradudin and A. R.  McGurn,  Physical Review E {\bf51}(6), 6301 (1995).
  
  
 \bibitem{Deych98} 
L. I. Deych, D. Zaslavsky and A. A.  Lisyansky,  Phys. Rev. Lett. {\bf81}(24), 5390 (1998).
  
  
  \bibitem{Han2008}
P. Han and C. Zheng,  Phys. Rev. E {\bf77}(4), 041111 (2008).
  
  
  \bibitem{Maurel2008}
A. Maurel and V. Pagneux, Phys. Rev. B {\bf78}(5), 052301 (2008).
  
  \bibitem{Maurel2010}
  A. Maurel, P. A. Martin and V. Pagneux, Waves in Random and Complex Media {\bf20}(4), 634-655 (2010).
  
  \bibitem{Izrailev2012} 
  F. M. Izrailev, A. A. Krokhin and N. M. Makarov, Physics Reports {\bf512}(3), 125-254 (2012).
  
  \bibitem{Maurel2013}
A.  Maurel and P. A. Martin, The European Physical Journal B {\bf86}(11), 1-10 (2013).
 
  \bibitem{Chang2003}
S. H. Chang, H. Cao and S. T. Ho, Quantum Electronics, IEEE Journal of Quantum Electronics {\bf39}(2), 364-374 (2003).

\bibitem{Croenne2011}
C. Croenne, E. J. S. Lee, H. Hu and J. H. Page, AIP Advances {\bf1}, 041401 (2011).

\bibitem{Xiao2011}
Y. Xiao, B. R. Mace, J. Wen and X. Wen, Phys. Lett. A {\bf375}, 1485 (2011).

\bibitem{Sugimoto1995}
N. Sugimoto and T. Horioka, J. Acoust. Soc. Am. {\bf97}(3), 1446 (1995). 

\bibitem{Bradley1994}
C. E. Bradley, J. Acoust. Soc. Am. {\bf96}(3), 1844 (1994).

\bibitem{Theocharis2014}
 G. Theocharis, O. Richoux, V. Romero-Garcia, A. Merkel and V. Tournat, Limits of slow sound propagation and transparency in lossy locally resonant periodic structures, accepted by
 New Journal of Physics (2014).
  
  \bibitem{Olivier} 
  O. Richoux and V. Pagneux, EPL {\bf59}(1), 34 (2002).
 
  \bibitem{Richoux2009} 
  O. Richoux, E. Morand and L. Simon, Annals of Physics, {\bf324}(9), 1983-1995 (2009).
  
  \bibitem{Richoux2007}
  O. Richoux, V. Tournat and T.  LeVanSuu, Phys. Rev. E, {\bf75}(2), 026615 (2007).

  \bibitem{MinutePub}
The sensor is developed jointly by CTTM (Centre de Transfert de Technologie du Mans, 20 rue Thal\`es de Milet, 72000 Le Mans, France) and LAUM (Laboratoire d'Acoustique de l'Universit\'e du Maine, UMR CNRS 6613, Avenue Olivier Messiaen, 72085 Le Mans Cedex 9, France) 
  
  \bibitem{Pierce}
 A.D. Pierce, Acoustics: An Introduction to Its Physical Principles and Applications, Acoustical Society of America (1989).
  
  \bibitem{Macaluso2011}
C. A.   Macaluso and J. P. Dalmont, J. Acoust. Soc. Am. {\bf129}(1), 404-414 (2011).
 
\end{thebibliography}
\end{document}